\documentclass[letterpaper]{article} 
\usepackage{aaai25}  
\usepackage{times}  
\usepackage{helvet}  
\usepackage{courier}  
\usepackage[hyphens]{url}  
\usepackage{graphicx} 
\usepackage{natbib}  
\usepackage{caption} 
\usepackage{amsmath}
\usepackage{multirow}
\usepackage{algorithm}
\usepackage{algorithmic}
\usepackage{comment}
\usepackage{graphicx}
\usepackage{makecell}
\usepackage{amssymb}
\usepackage{xcolor}
\usepackage{newfloat}
\usepackage{listings}
\usepackage{booktabs}
\DeclareCaptionStyle{ruled}{labelfont=normalfont,labelsep=colon,strut=off} 
\floatstyle{ruled}
\newfloat{listing}{tb}{lst}{}
\floatname{listing}{Listing}
\title{GASA-UNet: Global Axial Self-Attention U-Net for 3D Medical Image Segmentation}
\author{
    Chengkun Sun\textsuperscript{\rm 1},
    Russell Stevens Terry\textsuperscript{\rm 2},
    Jiang Bian\textsuperscript{\rm 1},
    Jie Xu\textsuperscript{\rm 1}
}
\affiliations {
    \textsuperscript{\rm 1}Department of Health Outcomes and Biomedical Informatics, University of Florida, Gainesville, FL 32611, USA\\
    \textsuperscript{\rm 2}Department of Urology, University of Florida, Gainesville, FL 32611, USA\\
    sun.chengkun@ufl.edu,
    bianjiang@ufl.edu,
    russell.terry@urology.ufl.edu,
    xujie@ufl.edu
}

\usepackage{bibentry}

\begin{document}
\nocopyright
\maketitle

\begin{abstract}
  Accurate segmentation of multiple organs and the differentiation of pathological tissues in medical imaging are crucial but challenging, especially for nuanced classifications and ambiguous organ boundaries. To tackle these challenges, we introduce GASA-UNet, a refined U-Net-like model featuring a novel Global Axial Self-Attention (GASA) block. This block processes image data as a 3D entity, with each 2D plane representing a different anatomical cross-section. Voxel features are defined within this spatial context, and a Multi-Head Self-Attention (MHSA) mechanism is utilized on extracted 1D patches to facilitate connections across these planes. Positional embeddings (PE) are incorporated into our attention framework, enriching voxel features with spatial context and enhancing tissue classification and organ edge delineation.  
  Our model has demonstrated promising improvements in segmentation performance, particularly for smaller anatomical structures, as evidenced by enhanced Dice scores and Normalized Surface Dice (NSD) on three benchmark datasets, i.e., BTCV, AMOS, and KiTS23.
\end{abstract}

%
\section{Introduction}
\label{sec:intro}
In recent decades, the ubiquity of clinical imaging technologies, such as X-ray, computed tomography (CT), and magnetic resonance imaging (MRI) has catalyzed extensive research in computational medical imaging analysis~\cite{shen2017deep, kasban2015comparative, xuxu}. 
Unlike natural images, medical images, though less semantically rich due to the absence of color, present more intricate spatial information pertinence to effective medical diagnosis and treatment planning. The challenge is accentuated when distinguishing between marginal tissues of different organs and interpreting ambiguous semantic information from pathological tissues~\cite{baumgartner2019phiseg, kohl2018probabilistic}. Moreover, the scarcity of medical images for deep learning research, compounded by the necessity to protect patient privacy and ensure high-quality, expertly labeled training data, restricts the size of datasets available compared to those for natural images~\cite{willemink2020preparing}. Although transfer learning from natural image domains is promising, the unique modalities and anatomical variations in medical images introduce a substantial gap that may diminish model performance~\cite{morid2021scoping}.

The seminal U-Net architecture~\cite{ronneberger2015u}, proposed in 2015, marked a significant advancement in biomedical image segmentation by fusing semantic information from both shallow and deep network layers to handle small-sample datasets. This model, with its encoder-decoder framework featuring skip connections, has demonstrated considerable success across various segmentation tasks and has inspired numerous model advancements and applications, even beyond the biomedical domain, (e.g., hyperspectral images~\cite{miao2019net}, image super-resolution~\cite{hu2019runet}, remote sensing imaging~\cite{wang2022unetformer}). Subsequently, nnUNet~\cite{isensee2021nnu} emerged as a leader for 3D medical segmentation with a modified 3D U-Net~\cite{cciccek20163d} structure, suggesting that a combination of meticulous image preprocessing and robust data augmentation, rather than model complexity, contributes significantly to superior segmentation results. This perspective underscores the utility of efficient processing pipelines over extensive parameter tuning, often achieving performance superior to conventional convolutional neural network (CNN)-based algorithms. Nonetheless, these approaches tend to extract spatial information through 3D convolutions in a generalized manner, which does not adequately address the detailed spatial interconnections between local and global features.

\begin{figure*}[htbp]
  \centering
   \includegraphics[width=0.8\textwidth]{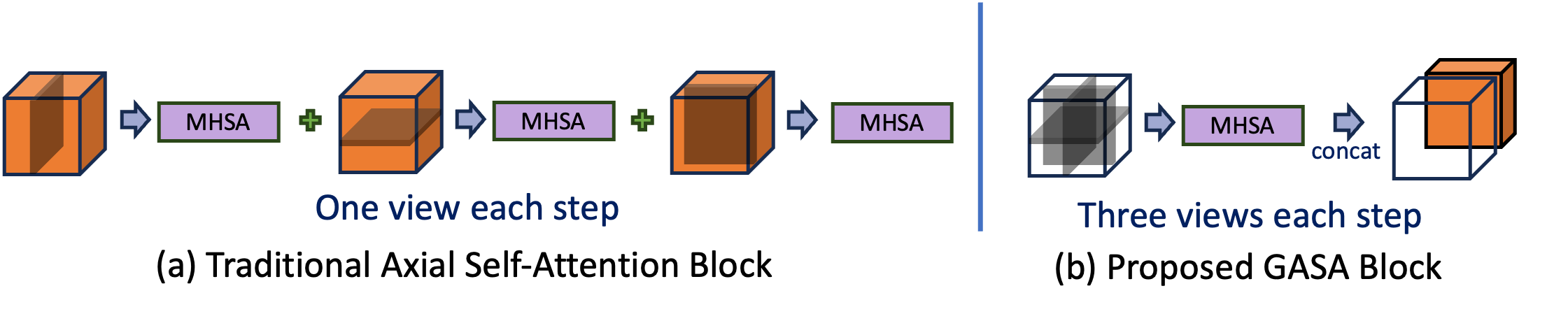}
   \caption{Comparison of Axial Self-Attention blocks: (a) Traditional Axial Self-Attention block, and (b) Proposed GASA block. }
   \label{fig:novelty}
\end{figure*}

On the other hand, in 2017, the Multi-Head Self-Attention (MHSA) mechanism~\cite{vaswani2017attention}, the key structure of the Transformer, was first proposed to perform translation tasks by introducing a global receptive field and positional awareness. This innovation set the stage for the Vision Transformer (ViT)~\cite{dosovitskiy2020image}, which applied the Transformer's encoder directly to image processing, adept at discerning long-range dependencies, which are typically out of reach for conventional small-kernel convolutional layers~\cite{han2022survey}. This advancement prompted investigations into the potential of ViT to augment the 3D U-Net framework. The prevailing approach has been to replace U-Net's encoder or decoder with the Transformer, aiming to exploit its superior global context modeling capabilities. However, this integration typically results in a substantial increase in the number of model parameters and necessitates pre-training on large datasets, which can be both complex and time-consuming. Furthermore, such direct substitution can diminish the model's ability to discern fine-grained local features due to the inherent design of ViTs, which prioritize global interactions over local ones. While positional embeddings (PE) within ViTs facilitate spatial relationships among image patches, they do not inherently provide a detailed, voxel-level spatial analysis, which is crucial for nuanced medical image segmentation tasks. 
As research progressed, axial attention, an offshoot of ViTs, emerged, applying a single axis of the tensor without flattening, effectively distilling high-dimensional data into compact and valuable traits~\cite{ho2019axial}. This approach's effectiveness in 2D imaging spatial distillation was demonstrated through Axial-deeplab~\cite{wang2020axial} and MedT~\cite{valanarasu2021medical}. Subsequently, studies extended axial information interaction to 3D medical image segmentation models~\cite{luu2021extending,liu20243d}. However, the conventional axial self-attention methodology typically employs three separate self-attention blocks for the computation of query (Q), key (K), and value (V) parameters along each axis (see Fig.~\ref{fig:novelty}(a)). This linear structure can lead to a gradual erosion of axial information initially captured by the attention mechanism. Subsequently, many medical image segmentation models (such as nnMamba~\cite{gong2024nnmamba}) have introduced the Mamba~\cite{gu2023mamba} architecture to improve performance. However, the performance of attention mechanisms in Mamba-based models still needs to be explored.

To utilize the global attention strengths of the ViT while incorporating voxel-level spatial details into 3D local features, we have developed a new component, termed the Global Axial Self-Attention (GASA) block. This block is integrated as an additional branch within the U-Net architecture, complementing the encoder and decoder without replacing them. In our design, the 3D feature space is mapped out as a coordinate system, with each distinct slice corresponding to a 2D plane. Voxel positions are pinpointed within this system. By adopting three orthogonal 2D convolution kernels in place of the traditional single 3D kernel, our approach generates patches from varied orientations, providing a comprehensive view. Subsequently, these patches undergo a self-attention process, coupled with channel concatenation and positional embedding, to numerically encode voxel locations. This preserves the foundational strength of U-Net in detailed local feature extraction while expanding its scope to include global spatial awareness, thus improving feature discernibility.

The contributions of this paper can be summarized in three key areas:
\vspace{-0.2cm}
\begin{itemize}
    \item Firstly, we present an innovative 3D self-attention block that merges global and local features within a revised vision transformer framework, endowing the U-Net model with 3D global axial self-attention capabilities with minimal parameter increase.
    \item Secondly, our model introduces a novel patch generation method that utilizes three 2D convolutional kernels, which improve the differentiation of semantically similar features through advanced spatial encoding. This includes expanding the output of self-attention values, concatenating channel dimensions, and employing positional embedding.
    \item Finally, comprehensive evaluations using BTCV, AMOS, and KiTS23 datasets reveal that our model achieves enhanced performance metrics in most instances, with only a nominal escalation in parameters when benchmarked against leading CNN-based, ViT-based, and Mamba-based models in the realm of medical imaging segmentation.
\end{itemize}

\begin{figure*}[h]
  \centering
   \includegraphics[width=1\textwidth]{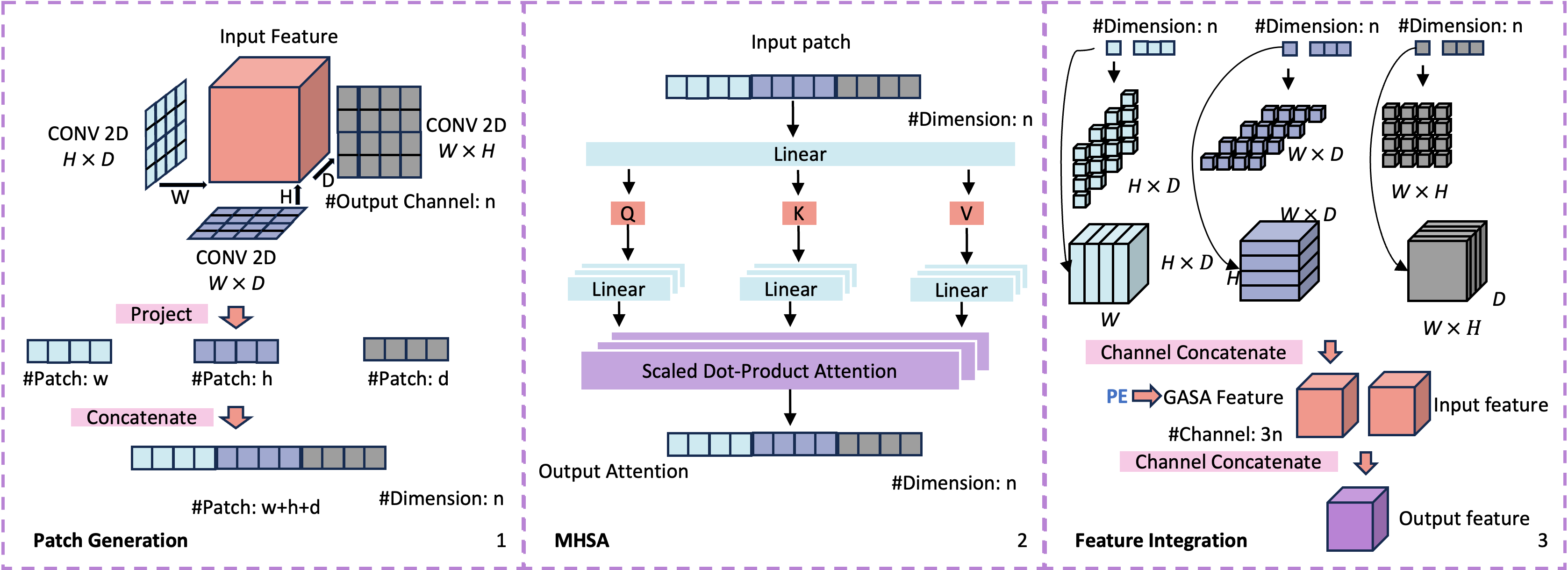}
   \caption{Schematic illustration of the GASA block. 
   The input features undergo three $2\times2$ convolutions along the W, H, and D axes. The resulting patches are concatenated and processed by the MHSA module with adjustable heads and dimensions. 
   Each output attention value is expanded into a specific slice and then into the entire 3D feature space. Following channel concatenation, these axial attention values are combined to form the GASA feature, which is further enhanced with absolute positional embedding. The input feature is integrated with the GASA feature at the channel level. Notably, the number of GASA feature channels increases from $n$ to $3n$ through a manual design process.
   }
   \label{fig:GASA}
\end{figure*}

\section{Related Work}
\subsection{Multi-Head Self-Attention}
MHSA~\cite{vaswani2017attention},  a fundamental component of the transformer architecture, operates by mapping a query to a set of key-value pairs. This process, essentially a weighted sum, forms dynamic connections among input patches, enabling the model to focus on different parts of the input for each prediction. The ViT~\cite{dosovitskiy2020image} has showcased that the application of transformer architectures to images can achieve outstanding performance in image classification tasks. With layers of MHSA and Multi-Layer Perceptron (MLP) blocks, ViT model leverages pre-training strategies akin to those used in natural language processing (NLP), proving to be effective even with datasets of limited size. The effectiveness of ViT in computer vision is largely ascribed to its ability to scale with training data, its embedded inductive biases, and the use of positional embeddings that capture and maintain essential positional information~\cite{dosovitskiy2020image}. Furthermore, the extensive parameterization inherent to transformer models is theorized to enhance the interpretability and effectiveness of ViTs suggesting that the large number of parameters may be instrumental in the models' success in vision-based tasks~\cite{han2022survey,neyshabur2018towards,livni2014computational}.

\subsection{Medical Vision Transformer}
Following U-Net's success in 3D medical imaging segmentation, numerous transformer based variants emerged. TransUNet~\cite{chen2021transunet} integrated transformers with CNN feature extractors, enhancing the extraction of fine-grained information, and empirical results indicated its superiority in leveraging self-attention over previous CNN-centric models. Subsequently, Swin-Unet~\cite{cao2022swin} 
set a new standard by replacing all convolutional layers with transformers, outperforming most full-convolution and hybrid architectures. The implementation of large-scale pre-trained models further bolstered transformers' efficiency. However, the increase in parameters and the labor-intensive pre-training demand substantial resources. The UNETR~\cite{hatamizadeh2022unetr} bypassed the need for pre-training by implementing 3D medical segmentation tasks with skip connections between the transformer encoder and CNN-based decoder. Furthermore, the Swinunetr~\cite{tang2022self} introduced novel pre-training methods, such as masked volume inpainting, image rotation, and contrastive coding, enhancing transformer performance. 

Despite these advances, achieving optimal performance without pre-trained models typically requires larger training datasets. Although the global receptive field provided by transformers improves feature extraction, the substitution of convolutions with ViTs can lead to a partial loss of local detail. Furthermore, transformer-based models tend to demand a high parameter count. Compared with CNN-based models of similar complexity, transformers do not always exhibit superior performance. Therefore, the application of transformers in 3D medical segmentation continues to be a promising area for further research and optimization.

\subsection{Axial Attention}
Axial attention~\cite{ho2019axial} was introduced as a computational strategy to manage the high demands of autoregressive models dealing with large-scale data. Unlike traditional methods that operate on flattened data structures, this technique applies self-attention along individual tensor axes, reducing computational demands. Axial-DeepLab~\cite{wang2020axial} introduced a position-sensitive axial-attention layer, outperforming conventional self-attention models in both efficiency and performance on ImageNet~\cite{russakovsky2015imagenet}. This improvement is achieved by factorizing 2D self-attention into two 1D attentions, thereby simplifying computational complexity and enabling modeling of spatial information over larger or global regions. Expanding on this approach, MedT~\cite{valanarasu2021medical} incorporated a gating mechanism within its axial attention module, yielding enhanced results in medical imaging. Building upon these advancements, subsequent studies by Luu et al.~\cite{luu2021extending} and Liu et al.~\cite{liu20243d} extended the application of 3D axial information in segmentation models, further refining the integration of axial attention in 3D medical imaging.

\section{GASA-UNet}
\label{GASA}
Our GASA-UNet model builds on the 3D encoder-decoder architecture of prevalent U-Net-like backbones by incorporating a novel GASA block.

\subsection{GASA Block}

The GASA block is the core component of the proposed model. It is designed to leverage the global attention strengths of the ViT~\cite{dosovitskiy2020image} while incorporating voxel-level spatial details into 3D local features. Fig.~\ref{fig:GASA} is the schematic illustration of the GASA block. 
Three $2\times2$ convolutions $F_{proj_W}$, $F_{proj_H}$, and $F_{proj_D}$ are performed on the input along the width (W), height (H), and depth (D) axes, each generating a 1D patch. These generated patches are then sequentially concatenated according to their respective W, H, and D axes. Specifically, the kernel size for the W axial direction is set to match the $H \times D$ dimensions of the global features. Correspondingly, for the H axial direction, the kernel spans the $W \times D$ dimensions, and for the D axial direction, it covers the $W \times H$ dimensions. The total number of patches generated is the collective sum of the dimensions across the $W$, $H$, and $D$. The input channel depth for these 2D convolutions mirrors that of the global features, while the output channel depth is preset, with a default value of 25—a parameter that will be further explored in our ablation studies. The mathematical representation for the patch generation process is as follows:
\begin{equation}
\begin{aligned}
P_W = F_{proj_W}(X), &\ \ \ P_W\subset \mathbb{R}^{w},  X\subset \mathbb{R}^{w\times h\times d},\\
P_H = F_{proj_H}(X), &\ \ \ P_H\subset \mathbb{R}^{h},  X\subset \mathbb{R}^{w\times h\times d},\\
P_D = F_{proj_D}(X),
&\ \ \ P_D\subset \mathbb{R}^{d},  X\subset \mathbb{R}^{w\times h\times d},\\
P = [P_W,P_H,P_D], &\ \ \ P\subset \mathbb{R}^{w+h+d},
\end{aligned}
\end{equation}
where $X$ denotes the input 3D features, $P_W$, $P_H$, and $P_D$ represent the patches generated along the W, H, and D axes, respectively. $P$ stands for the aggregated patches. 
Here, $w$, $h$, and $d$ indicate the number of slices along the W-axis, H-axis, and D-axis, respectively.

Following patch generation, these patches are input directly into an MHSA block to compute the GASA attention values. This particular MHSA block is adapted from the ViT architecture but simplifies the structure by omitting the MLP layers typically included. The attention mechanism within this block processes the input patches, enabling the model to focus on different parts of the image and extract pertinent features. The GASA attention values are calculated with the function~\cite{vaswani2017attention,bahdanau2014neural,britz2017massive,press2016using} as follows: 
\begin{align}
Attention(Q, K, V ) = softmax(\frac{QK^{T} }{\sqrt{d_{k} } } )V,
\end{align}
where the $Q$, $K$, and $V$ correspond to the query, keys, and values respectively. $d{_k}$ represents the dimension of the keys. Notably, the output attention dimension remains consistent throughout the process.

Following this,  each axial attention output was expanded into a 2D feature, aligned with the direction of the convolution that produced it. Specifically, W-axis attention was reshaped to the $H \times D$ dimensions of the input features, while H and D-axis attentions matched the $W \times D$ and $W \times H$ dimensions, respectively. These axial features were then concatenated along the channel dimension, tripling the output dimensions compared to the MHSA output. A 1D learnable absolute positional embedding was integrated into each voxel of the GASA output across all channels, embedding comprehensive global information into the features. Finally, these GASA features were merged with the initial global features along the channel direction before being channeled into the CNN-based decoder, as depicted in Figure~\ref{fig:GASA}.

\subsection{GASA-UNet}

The GASA block, positioned at the end of the encoder and the beginning of the 3D CNN decoder, is a key addition to the U-Net-like backbone, facilitating axial attention. It encodes 3D spatial position information into the feature representation by processing slices in the W, H, and D directions of the feature space. This enables the extraction of axial attention in the three dimensions. 

Our loss function is a composition of soft Dice loss~\cite{milletari2016v} and robust cross-entropy loss, with an individual weight of 1, from the nnUNet. The loss function~\cite{hatamizadeh2022unetr} is formulated as follows:
\begin{equation}
\begin{aligned}
L &= 1 - \frac{1}{N_c}\sum_{i=1}^{N_c} \frac{2 \cdot \sum_{j=1}^{N_y} L_{i,j}Y_{i,j}}{\sum_{j=1}^{N_y} L_{i,j}^2 + \sum_{j=1}^{N_y} Y_{i,j}^2} \notag \\
& -\frac{1}{N_y}\sum_{i=1}^{N_c}\sum_{j=1}^{N_y} L_{i,j} \cdot \log(Y_{i,j}),
\end{aligned}
\end{equation}
where $N_y$ stands for the total number of voxels, and $N_c$ indicates the number of distinct classes. $L_{i,j}$, $Y_{i,j}$ represent the one-hot encoded label and the predicted probability for the $i$-th class and the $j$-th voxel. 

\begin{table*}[htbp]
  \centering
    \begin{tabular}{l|ll|ll|ll}
    \hline
     & \multicolumn{2}{c}{\textbf{BTCV}} & \multicolumn{2}{|c|}{\textbf{AMOS}}& \multicolumn{2}{c}{\textbf{KiTS23}}\\
    \cline{2-7}
    Models &Dice  & NSD   & Dice  & NSD &Dice  & NSD\\
    \hline
    nnMamba 
    & 75.47 & 67.05 & 83.16 & 73.00 &64.39&52.55\\
    \footnotesize{\ +GASA(Ours)} & 76.27\textcolor{red}{
    \footnotesize{ \ (+0.80)}} & 67.52\textcolor{red}{     \footnotesize{ \ (+0.47)}} & 84.31\textcolor{red}{     \footnotesize{ \ (+1.15)}} & 75.17\textcolor{red}{     \footnotesize{ \ (+2.17)}}&68.25\textcolor{red}{     \footnotesize{ \ (+3.86)}}&56.01\textcolor{red}{     \footnotesize{ \ (+3.46)}} \\
    \hline
    UNETR 
    & 69.65 & 59.18 & 68.62 & 51.65&59.38&46.28 \\

    \footnotesize{\ +GASA(Ours)} & 70.04\textcolor{red}{     \footnotesize{ \ (+0.39)}} & 59.87\textcolor{red}{     \footnotesize{ \ (+0.69)}} & 71.55\textcolor{red}{     \footnotesize{ \ (+2.93)}} & 55.00\textcolor{red}{     \footnotesize{ \ (+3.35)}} &59.87\textcolor{red}{     \footnotesize{ \ (+0.49)}}&47.57\textcolor{red}{     \footnotesize{ \ (+1.29)}}\\
    \hline
    V-Net 
    & 78.32 & 70.77 & 77.15 & 62.97 &62.62&50.47\\
    \footnotesize{\ +GASA(Ours)} & 78.47\textcolor{red}{     \footnotesize{ \ (+0.15)}} & 70.59\textcolor{red}{     \footnotesize{ \ (-0.18)}} & 79.12\textcolor{red}{     \footnotesize{ \ (+1.97)}} & 66.16\textcolor{red}{     \footnotesize{ \ (+3.19)}}&65.51\textcolor{red}{     \footnotesize{ \ (+2.89)}}&53.27\textcolor{red}{     \footnotesize{ \ (+2.8)}} \\
    \hline
    nnUNet
    & 81.52 & 76.10  & 89.75 & 85.58 &74.99&64.83\\

    \footnotesize{\ +GASA(Ours)} & \underline{81.69}\textcolor{red}{     \footnotesize{ \ (+0.17)}} & \underline{76.03}\textcolor{red}{     \footnotesize{ \ (-0.07)}} & \underline{89.83}\textcolor{red}{     \footnotesize{ \ (+0.08)}} & \underline{85.64}\textcolor{red}{     \footnotesize{ \ (+0.06)}}&\underline{75.97}\textcolor{red}{     \footnotesize{ \ (+0.98)}}&\underline{66.05}\textcolor{red}{     \footnotesize{ \ (+1.22)}} \\
    \footnotesize{\ +GASA-L(Ours)} & \textbf{82.41\textcolor{red}{     \footnotesize{ \ (+0.89)}}} & \textbf{77.04\textcolor{red}{     \footnotesize{ \ (+0.94)}}} & \textbf{89.87\textcolor{red}{     \footnotesize{ \ (+0.12)}}} & \textbf{85.93\textcolor{red}{     \footnotesize{ \ (+0.35)}}}&\textbf{76.49\textcolor{red}{     \footnotesize{ \ (+1.50)}}}&\textbf{67.21\textcolor{red}{     \footnotesize{ \ (+2.38)}}} \\
    \hline
    \end{tabular}%
    \caption{Comparison of segmentation performance in diverse backbones on the BTCV, AMOS, and KiTS23 dataset.}
    \vspace{-0.6cm}
  \label{tab:btcvamos}%
\end{table*}%

\section{Experiments}
\label{Experiment}
\subsection{Datasets}
We evaluated the proposed GASA-UNet model using several well-established medical imaging segmentation datasets:
\begin{itemize}
    \item \textbf{BTCV}: The Beyond the Cranial Vault (BTCV) abdomen challenge dataset~\cite{BTCV2015} comprises 30 public abdominal CT scans obtained during the portal venous contrast phase. 
    
    \item \textbf{AMOS}: Multi-Modality Abdominal Multi-Organ Segmentation (AMOS) 2022~\cite{ji2022amos} contains two types of imaging, CT and MRI, aimed at comprehensively evaluating the performance of an extended cross-modality CT and  MRI segmentation task. It includes 300 CT scans annotated for 15 organs and 60 MRI scans annotated for 13 organs. 
    
    \item \textbf{KiTS23}: The 2023 Kidney and Kidney Tumor Segmentation Dataset~\cite{heller2023kits21} (KiTS23), building on KiTS19 and the KiTS21, offers a training set of 489 cases from the nephrogenic contrast phase or late arterial phase. The dataset has maintained and expanded upon a public cohort of hundreds of CT scans, along with semantic segmentation of kidneys, renal tumors, and renal cysts. 
\end{itemize}

\subsection{Experimental Setup}
\textbf{Baselines.} We incorporated our proposed GASA-UNet model into several backbones. These include nnUNet~\cite{isensee2021nnu}, which serves as a baseline due to its robust performance across various medical imaging tasks. We also benchmarked against UNETR~\cite{hatamizadeh2022unetr} which is known for its transformer-based architecture tailored for medical imaging, V-Net~\cite{milletari2016v} which is a typical CNN-based model, nnMamba~\cite{gong2024nnmamba} which is the latest Mamba-based segmentation model.

Furthermore, we incorporated SOTA 3D axial attention blocks in the nnUNet, including conventional axial attention (AT)~\cite{luu2021extending,wang2020axial,ho2019axial} and medical axial transformer (MAT)~\cite{liu20243d,valanarasu2021medical}, along with the spatial and channel attention block CBAM~\cite{woo2018cbam}, all integrated at the location where the GASA block was implemented. We skipped the experiments using MAT on the AMOS dataset because MAT cannot process asymmetric 3D features.

\textbf{Implementation.} Our implementation adheres closely to the nnUNet framework~\cite{isensee2021nnu}, from data preprocessing and augmentation to model training and inference. The scans and labels were resampled to the same spacing according to nnUNet. In pursuit of an end-to-end model, we omitted nnUNet's post-processing steps to focus on the model's intrinsic segmentation performance. To investigate the GASA block's adaptability to larger structures, we expanded the nnUNet encoder's scale by integrating three ResNet~\cite{he2016deep} blocks after each encoder layer, except for the final layer, where we incorporated five ResNet blocks. This configuration is showcased in the large GASA model (referred to as GASA-L in the experiments section). By positioning these modifications at the encoder's terminal point, we have observed an enhancement in the GASA block's efficacy. 
For testing, we utilized the model from the final training epoch. Training was conducted on an Nvidia A100 GPU with 40GB of memory, with each training fold taking approximately 20 hours.

For a fair comparison, we preserved the default configurations when reproducing nnUNet, UNETR, 
 V-Net, and nnMamba, we aligned them with nnUNet settings. More implementation details can be found in supplementary materials.

\textbf{Evaluation.} 
For the BTCV dataset, we allocated 12 scans to the test set and 18 scans to the training and validation set. For the AMOS dataset, 320 scans were divided into 240 for training and validation and 80 for testing, maintaining a training-to-validation ratio of 4:1. We conducted 5-fold cross-validation on all models, averaging their softmax output across folds to determine voxel probabilities. For the KiTS23 dataset, 89 CT scans were randomly assigned for testing, with the remaining scans used for training and validation.

Our evaluation relies on the Dice score~\cite{milletari2016v} and Normalized Surface Dice (NSD)~\cite{nikolov2018deep} metrics for both the BTCV and AMOS datasets. For the KiTS23 dataset, which features CT scans with variable contrasts and segmentation targets closely resembling real-life scenarios, we adopted the ``Hierarchical Evaluation Classes'' (HECs) as the evaluation entities, in accordance with the KiTS challenge guidelines~\cite{heller2023kits21}. HECs group subcategories into their parent categories for metric calculation, enabling a more detailed assessment. In KiTS23, this includes categories like kidney and masses (kidney, tumor, cyst), kidney mass (tumor, cyst), and tumor only, with Dice scores and NSD calculated to reflect these hierarchical groupings. 

\subsection{Segmentation Results}
\label{result}

\textbf{Backbones Comparison}.
Tables~\ref{tab:btcvamos} present the Dice scores and NSD performances on the BTCV, AMOS, and KiTS23 datasets, respectively based on different backbones. Our nnUNet-GASA-L model performed the best. When integrating the GASA module into nnUNet (CNN-based), V-Net (CNN-based), UNETR (Transformer-based), and nnMamba (Mamba-based) architectures, there was a general improvement in baseline performance. Specifically, on the BTCV, AMOS, and KiTS23 datasets, we observed Dice score increases from 0.15 to 0.8, 1.15 to 1.97, and 0.49 to 3.86, respectively. Additionally, there were improvements in NSD from 0.47 to 0.69 (with a 0.18 decrease in V-Net), 2.17 to 3.35, and 1.29 to 3.46, respectively. The nnUNet with GASA-L achieved the best performance, showcasing superior generalization and feature extraction capabilities despite the small training set size and the absence of sophisticated lesion tissue segmentation targets.

\begin{table*}[htbp]
  \centering
    \begin{tabular}{l|ll|ll|ll}
    \hline
     & \multicolumn{2}{c|}{\textbf{BTCV}} & \multicolumn{2}{c|}{\textbf{AMOS}}& \multicolumn{2}{c}{\textbf{KiTS23}}\\
    \cline{2-7}
    Models &Dice  & NSD   & Dice  & NSD &Dice  & NSD\\
    \hline
    nnUNet
    & 81.52 & 76.10  & 89.75 & 85.58 &74.99&64.83\\

    +CBAM
    & 81.48\textcolor{blue}{\footnotesize{ \ (-0.04)}} & 75.84\textcolor{blue}{\footnotesize{ \ (-0.26)}} & 89.81\textcolor{blue}{\footnotesize{ \ (+0.06)}} & 85.58\textcolor{blue}{\footnotesize{ \ (+0)}} &74.88\textcolor{blue}{\footnotesize{ \ (-0.11)}}&64.85\textcolor{blue}{\footnotesize{ \ (+0.02)}}\\

    +AT 
    & 81.35\textcolor{blue}{\footnotesize{ \ (-0.17)}} & 75.96\textcolor{blue}{\footnotesize{ \ (-0.14)}} & 89.62\textcolor{blue}{\footnotesize{ \ (-0.13)}} & 85.56\textcolor{blue}{\footnotesize{ \ (-0.02)}} &75.65\textcolor{blue}{\footnotesize{ \ (+0.66)}}&65.57\textcolor{blue}{\footnotesize{ \ (+0.74)}}\\

    +MAT 
    & 81.29\textcolor{blue}{\footnotesize{ \ (-0.23)}} & 75.83\textcolor{blue}{\footnotesize{ \ (-0.27)}} & - & - &75.27\textcolor{blue}{\footnotesize{ \ (+0.28)}}&65.10\textcolor{blue}{\footnotesize{ \ (+0.27)}}\\

    +GASA(Ours) & \underline{81.69}\textcolor{red}{     \footnotesize{ \ (+0.17)}} & \underline{76.03}\textcolor{red}{     \footnotesize{ \ (-0.07)}} & \underline{89.83}\textcolor{red}{     \footnotesize{ \ (+0.08)}} & \underline{85.64}\textcolor{red}{     \footnotesize{ \ (+0.06)}}&\underline{75.97}\textcolor{red}{     \footnotesize{ \ (+0.98)}}&\underline{66.05}\textcolor{red}{     \footnotesize{ \ (+1.22)}} \\
    +GASA-L(ours) & \textbf{82.41\textcolor{red}{     \footnotesize{ \ (+0.89)}}} & \textbf{77.04\textcolor{red}{     \footnotesize{ \ (+0.94)}}} & \textbf{89.87\textcolor{red}{     \footnotesize{ \ (+0.12)}}} & \textbf{85.93\textcolor{red}{     \footnotesize{ \ (+0.35)}}}&\textbf{76.49\textcolor{red}{     \footnotesize{ \ (+1.50)}}}&\textbf{67.21\textcolor{red}{     \footnotesize{ \ (+2.38)}}} \\
    \hline
    \end{tabular}%
    \caption{Comparison of segmentation performance with different axial attentions on the BTCV, AMOS and KiTS23 dataset.}
  \label{tab:axialattention}%
\end{table*}%

\begin{figure*}[!htbp]
  \centering
   \includegraphics[width=0.95\textwidth]{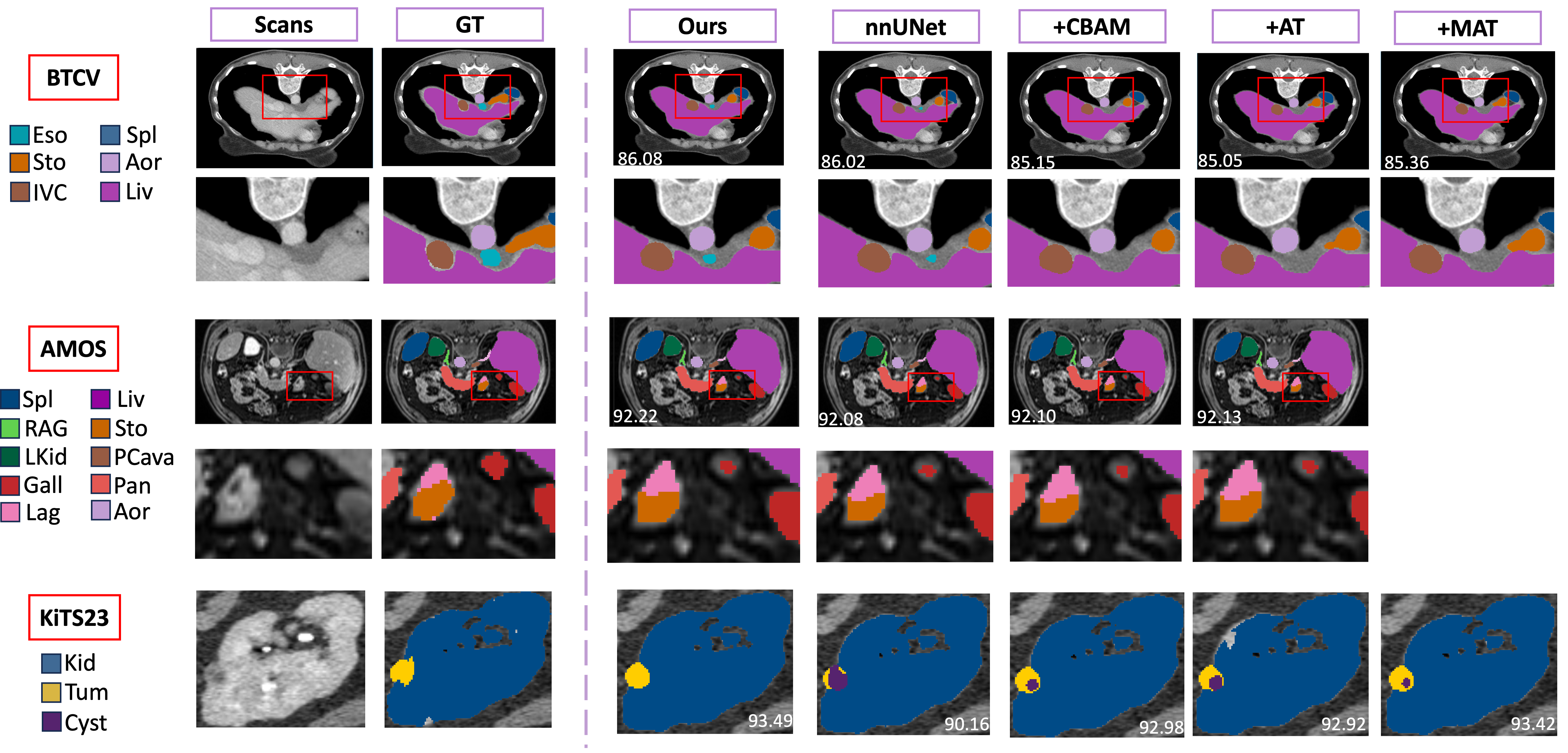}
   \caption{Visual comparison of segmentation results by different axial attentions on representative samples from BTCV, AMOS, and KiTS23 datasets. 
   The +MAT results on AMOS were omitted due to its inability to process asymmetric 3D features.}
   \label{fig:outcome_att}
\end{figure*}

\begin{table*}[!htbp]
  \centering
  \resizebox{\textwidth}{!}{
 \begin{tabular}{@{}l|cccc|cccccc|@{}cccc}
    \cline{1-9}\cline{11-15} 
    Head/Dim&1/50&5/50&10/50&25/50&2/10&5/25&10/50&20/100&&PE&nnUNet&No\_PE&B\_PE&A\_PE\\
    \cline{1-9}\cline{11-15} 
    BTCV (Avg)&81.65&81.54&81.69&81.58&\underline{81.23}&81.64&81.69&81.50&& &\underline{81.51}&81.44&81.13&\textbf{81.63}\\
    \cline{1-9}\cline{12-15} 
    AMOS (Avg)&89.75&89.83&89.79&89.74&\underline{87.99}&89.82&89.79&89.79&&&\underline{89.75}&89.79&89.59&\textbf{89.82}\\
    \cline{1-9}\cline{12-15} 
    KiTS23 (Avg)&\underline{74.61}&75.97&75.14&75.96&75.28&75.86&75.14&75.99&&&\underline{74.99}&75.29&75.43&\textbf{75.86}\\
    \cline{1-9}\cline{11-15} 
  \end{tabular}
  }
  \caption{Impact of Head/Dimension (Dim) configuration and PE on performance.}
  \label{tab:abla}
\end{table*}

\begin{figure*}[!htbp]
  \centering
   \includegraphics[width=0.95\textwidth]{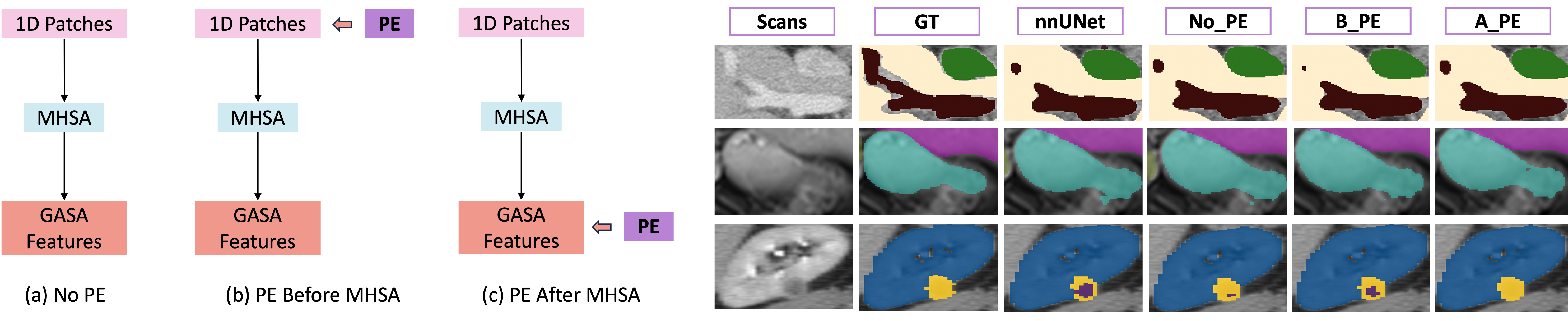}
   \captionof{figure}{Overview of the three PE strategies and visual comparison on selected samples using three PE strategies.}
   \label{fig:PE_strategy}
\end{figure*}

\textbf{Axial Attentions Comparison}.
 Tables~\ref{tab:axialattention} present the Dice scores and NSD performances on the BTCV, AMOS, and KiTS23 datasets based on different axial attentions, respectively. On the BTCV dataset, the GASA block achieved an average improvement of 0.2\% to 0.4\% in Dice Score and 0.07\% to 0.2\% in NSD compared to other attention mechanisms. On the AMOS dataset, our model achieved a slight improvement from 0.02\% to 0.21\% in Dice Score and from 0.06\% to 0.08\% increase in NSD, underscoring its proficiency in delineating organ margins. Validated on a multi-modality dataset without specialized complex lesion segmentation, these results confirm our model's enhanced generalization and multi-organ feature extraction capabilities. On the KiTS23 dataset, our model modestly exceeded nnUNet's performance, indicating an improved ability to segment pathological tissues within complex structures like tumors. By incorporating the GASA Block, our axial attention mechanism outperformed other axial attentions by about 0.32\% to 1.09\% in Dice Score and 0.48\% to 1.2\% in NSD.

\subsection{Ablation Experiments}
In our ablation studies, we aimed to determine the optimal configuration of hidden layer dimensions and heads across all datasets upon small GASA-Unet. Initially, we maintained a fixed ratio of 5:1 for dimensions to heads, testing dimensions of 10, 25, 50, and 100. Subsequently, we examined the impact of varying the number of heads, i.e., 1, 5, 10, 25, while keeping the dimensions at 50 within the MHSA to assess their influence on performance. The evaluations were based on average dice scores. Our findings, as shown in Table~\ref{tab:abla}, indicate that a GASA configuration with two heads and 10 dimensions underperformed by approximately 2.7\% in AMOS and by 0.2-0.35 percent in BTCV. Similarly, using 1 head and 50 dimensions was the least effective on KiTS23. Optimal performance requires more than 1 head and more than 10 dimensions, but beyond a certain point, increased numbers do not yield further gains. Thus, we opted for 5 heads and 25 dimensions for our GASA block.

We also assessed the impact of positional embedding (PE) using the model with 25 dimensions and 5 heads. We implemented two positional embedding strategies, as shown in Figure~\ref{fig:PE_strategy}. The first followed the Vision Transformer approach by adding a learnable 1D absolute positional embedding post-patch extraction. The second strategy involved adding a learnable 1D absolute positional embedding to the GASA features to anchor the 3D coordinates for each voxel. We also gauged performance without positional embedding for comparison. As indicated in Table~\ref{tab:abla}, incorporating a 1D learnable positional embedding subsequent to the MHSA yielded the best GASA model, producing enhanced axial self-attention with an increase in dice scores by 0.12 percent on BTCV, 0.07 percent on AMOS, and 0.87 percent on KiTS when compared to the baseline nnUNet model.

\subsection{Computational Load}
We analyzed the computational load, measuring both the number of floating-point operations per second (FLOPs) and the parameter count for each model tested on the BTCV dataset. Our default configuration for the GASA block entails 5 heads and 25 dimensions. As outlined in Table~\ref{tab:para}, incorporating the GASA block into the nnUNet framework marginally increases the parameters by approximately 1M and the FLOPs by around 0.3G. This increase is justified by the improved capability of the model to process spatial information and distinguish tissues with indistinct boundaries, without a significant addition to the computational burden.

\begin{table}[htbp]
  \centering
  \resizebox{8.5cm}{!}{
    \begin{tabular}{c|c|c|c|c|c}
    \hline
    Models  & nnUNet &+CBAM & +AT & +MAT & +GASA\\
    \hline
    Params (M) & 29.54 & 29.57 & 29.54 & 30.35 & 30.55 \\
    \hline
    FLOPs (G) & 662.9 & 662.9 & 662.9 & 662.9 & 663.2 \\
    \hline
    \end{tabular}
    }
    \caption{Comparative analysis of computational load: parameter count and FLOPs.}
  \label{tab:para}
\end{table}

\section{Discussion}
Our GASA-UNet model achieved top results across BTCV, AMOS, and KiTS23 datasets using a 5-fold validation approach, notably without pre-training or post-processing strategies. The GASA block's refined ability to distinguish similar pathological tissues from healthy ones, and to clarify blurred tissue across different organs, was evident. Yet, we must acknowledge some limitations. One primary challenge is the model's current sensitivity to tissues at organ margins, where indistinct tissues are common. Our use of dice loss and soft cross-entropy loss may not adequately emphasize these marginal areas, suggesting a need to explore loss functions that focus on edge delineation. Additionally, the variability inherent in biological tissue annotation, particularly at margins, can constrain the GASA block's potential for precise voxel-level segmentation. This natural variability in annotations presents a hurdle that we have yet to fully overcome. 

Our integration of the modified Vision Transformer was limited to a central component of the U-Net architecture; its application across all convolutional layers remains untested and could potentially open new avenues for model enhancement. Furthermore, while our ablation studies focused on the placement of 1D learnable positional embedding and its absence, other embedding techniques could offer further improvements and warrant exploration. Although our model improved without pre-trained models, the full extent of pre-training's benefits is still to be thoroughly assessed. Our model's foundation is derived from the nnUNet framework, retaining all its default settings. Hyperparameter tuning could potentially elevate performance, despite being beyond our current scope. Moreover, the model's performance could vary across different GPU architectures. Lastly, the GASA block's applicability to other models for 3D tasks beyond segmentation, like reconstruction and recognition, also remains an area ripe for exploration.

\section{Conclusion}
\label{discussion}
In this study, we introduced an innovative framework named GASA-UNet that integrated 3D spatial perspectives into medical image analysis. By employing 2D convolutional projections rather than 3D approaches, we effectively generated patches enriched with 3D axial information, thereby reducing model complexity. The integration of MHSA and channel concatenation techniques allowed us to infuse spatial correlations into voxel features derived from convolutions. This compact 3D block has proven effective in clarifying similar voxel features, leading to a modest improvement in segmentation performance on the BTCV, AMOS, and KiTS23 datasets.

\bibliography{aaai25}

\end{document}


\twocolumn[
\begin{center}
    \section*{Technical Appendix}  
\end{center}  
]
\section{Implementation Details}
\label{sec:Implementation details}

The standard 3D nnUNet~\cite{isensee2021nnu} was employed to validate axial attention blocks. We adhered to the nnUNet's default development schedule for nnMamba, UNETR, and V-Net.

For image normalization, we clipped the intensity values using the 0.5th and 99.5th percentiles of foreground voxels and normalized using the global foreground mean and standard deviation. Image resampling was carried out using third-order spline interpolation. In cases of anisotropic images, in-plane resampling was performed with third-order spline interpolation, and out-of-plane resampling was done using nearest-neighbor interpolation. Segmentation maps were resampled by converting them to one-hot encodings, interpolating each channel with linear interpolation, and then obtaining the segmentation mask through an argmax operation. In anisotropic cases, the low-resolution axis was interpolated using the nearest-neighbor approach. The target spacing for the lowest resolution axis was set to the tenth percentile of spacings found in the training cases, applied if both voxel and spacing anisotropy (i.e., the ratio of the lowest spacing axis to the highest spacing axis) exceeded a factor of three.

Networks were trained using a batch size of 2, over 1,000 epochs, with 250 iterations per epoch, and a fixed seed. We utilized stochastic gradient descent with Nesterov momentum, setting the momentum coefficient ($\mu$) to 0.99. The initial learning rate was set at 0.01, and followed a ``poly'' decay policy~\cite{chen2017deeplab}, as defined by $(1 - \text{epoch}/\text{epoch}_{\max})^{0.9}$. Specifically, for the model nnMamba, the original learning rate of 0.0001 from the paper was preserved. During inference, a sliding window approach was adopted, where the window dimensions matched the training phase patch size. Overlapping predictions by 50\% and applying Gaussian importance weighting enhanced the predictions. Additionally, test-time augmentation (TTA) was performed by mirroring along all axes.


The GASA Block, when calculating QKV (Query, Key, Value), applies layer normalization~\cite{ba2016layer} after each fully connected layer, and after obtaining the attention values, a 0.5 Dropout is applied. This configuration is typically used to improve model training stability and prevent overfitting.
Specifically, we found that when incorporating this model into nnUNet, removing layer normalization yielded better results. After removing layer normalization, the model's Dice score on the KiTS dataset improved from 75.12 to 75.99.
\newpage
\section{Experiment Results}
We provide the segmentation results for each organ or tissue on the BTCV, AMOS, and KiTS datasets, with precise DICE and NSD scores.
\begin{table}[h]
  \centering
  \resizebox{0.45\textwidth}{!}{
  \begin{tabular}{@{}c|cc@{\hspace{1em}}cc@{\hspace{1em}}cc@{\hspace{1em}}cc@{}}
    \hline
    &\multicolumn{2}{c}{Kid \& Masses} & \multicolumn{2}{c}{Mass}& \multicolumn{2}{c}{Tumor} &\multicolumn{2}{c}{Avg} \\
    \hline
    Model&Dice&NSD&Dice&NSD&Dice&NSD&Dice&NSD\\
    \hline
    nnMamba&80.54&69.16&63.63&51.35&49.00&37.14&64.39&52.55\\
    \hline
    +GASA(Ours)&86.99&75.76&66.92&54.07&50.84&38.21&68.25&56.01\\
    \hline    
    \hline
    UNETR&73.01&62.92&55.33&40.51&49.78&35.40&59.38&46.28\\
    \hline
    +GASA(Ours)&81.73&72.26&51.98&38.04&45.90&32.42&59.87&47.57\\
    \hline
    \hline

    V-Net&82.80&72.33&57.82&44.25&47.23&34.82&62.62&50.47\\
    \hline
    +GASA(Ours)&86.52&75.91&59.75&46.29&50.26&37.63&65.51&53.27\\
    \hline
    \hline
    nnUNet&89.94&84.04&72.79&60.49&62.23&49.96&74.99&64.83 \\
    \hline
    +CBAM&91.02&85.23&71.82&59.80&61.79&49.53&74.88&64.85\\
    \hline
    +AT&91.01&85.33&\underline{73.17}&60.59&62.78&50.77&75.65&65.57\\
    \hline
    +MAT&89.76&83.89&72.75&60.53&\underline{63.31}&50.90&75.27&65.10\\
    \hline
    +GASA(Ours)&\underline{91.21}&\underline{85.65}&73.06&\underline{60.68}&\textbf{63.63}&\textbf{51.81}&\underline{75.97}&\underline{66.05} \\
    \hline
    +GASA-L(Ours)&\textbf{92.76}&\textbf{87.11}&\textbf{74.23}&\textbf{62.75}&62.48&\underline{51.77}&\textbf{76.49}&\textbf{67.21} \\
    \hline

  \end{tabular}}
  \caption{Comparison of segmentation performance on the KiTS23 dataset.} 
  \label{tab:KiTS}
\end{table}

\begin{table*}
  \centering
  \resizebox{0.9\textwidth}{!}{
  \begin{tabular}{@{}c|ccccccccccccc|c@{}}
    \hline
    \textbf{Dice}&Spl&Rkid&Lkid&Gall&Eso&Liv&Sto&Aor&IVC&Vei&Pan&Rag&Lag&Avg  \\

    
    \hline
    nnMamba&85.11&88.35&87.43&49.13&73.91&92.88&77.32&89.83&84.90&64.07&64.40&59.12&64.67&75.47 \\
    \hline
    +GASA(Ours)&84.27&89.53&87.71&52.27&75.75&93.16&78.23&89.76&84.06&63.09&67.27&61.47&64.88&76.27 \\
    \hline
    \hline
    UNETR&80.41&78.01&76.86&51.50&68.72&88.40&63.68&87.01&81.56&59.23&55.49&60.63&53.93&69.65 \\
    \hline
    +GASA(Ours)&81.15&77.26&76.83&51.84&67.86&88.38&64.04&86.98&81.66&60.39&58.79&61.32&53.96&70.04 \\
    \hline
    \hline
    V-Net&85.18&83.27&83.42&62.33&77.27&92.88&80.89&89.40&85.76&67.18&74.15&67.56&68.87&78.32 \\
    \hline
    +GASA(Ours)&85.26&83.57&85.12&64.85&76.89&92.80&82.44&88.52&85.46&67.50&73.79&67.58&66.36&78.47 \\
    \hline
    \hline
    nnUNet&\underline{86.54}&87.27&84.13&68.47&79.09&93.72&\underline{85.38}&\textbf{93.05}&86.89&\underline{71.80}&\underline{82.09}&\underline{71.11}&\textbf{72.79}&81.52 \\
    \hline
    +CBAM&\textbf{86.59}&\underline{89.71}&84.81&67.48&79.34&\textbf{93.75}&84.20&90.76&86.81&71.55&81.88&70.05&72.29&81.48 \\
    \hline
    +AT&86.50&86.16&84.07&68.17&\underline{79.47}&93.72&84.50&91.01&\underline{87.29}&\underline{71.80}&81.89&70.36&\underline{72.66}&81.35 \\
    \hline
    +MAT&86.42&89.69&85.01&64.91&79.01&93.69&84.71&90.63&87.22&71.20&81.49&70.47&72.27&81.29 \\
    \hline
    +GASA(Ours) &86.41&\textbf{90.21}&\underline{85.44}&\underline{69.29}&78.80&\underline{93.74}&84.42&90.50&86.94&71.56&81.73&70.98&71.98&\underline{81.69} \\
    \hline
    +GASA-L(Ours)&86.30&88.85&\textbf{86.01}&\textbf{72.78}&\textbf{79.51}&\textbf{93.75}&\textbf{86.60}&\underline{91.21}&\textbf{87.42}&\textbf{72.19}&\textbf{83.08}&\textbf{71.49}&72.11&\textbf{82.41} \\
    \hline
    \hline
    \textbf{NSD}&Spl&Rkid&Lkid&Gall&Eso&Liv&Sto&Aor&IVC&Vei&Pan&Rag&Lag&Avg  \\
    \hline

    
    \hline
    nnMamba&76.90&81.64&78.59&31.15&67.52&77.65&53.73&83.46&72.45&60.94&48.69&67.77&71.13&67.05 \\
    \hline
    +GASA(Ours)&76.70&81.45&78.89&34.70&69.76&77.95&53.45&82.54&71.20&59.46&50.74&69.69&71.27&67.52 \\
    \hline
    \hline
    UNETR&67.85&69.22&65.25&37.70&62.82&69.89&33.94&79.35&64.80&53.88&39.70&67.61&57.27&59.18 \\
    \hline
    +GASA(Ours)&68.31&69.39&67.01&38.58&62.60&69.09&35.06&79.12&65.29&54.93&42.46&69.02&57.54&59.87 \\
    \hline
    \hline
    V-Net&77.23&77.91&76.82&50.90&70.65&76.76&57.91&84.36&73.18&63.80&61.07&64.63&74.79&70.77 \\
    \hline
    +GASA(Ours)&77.73&78.31&78.63&50.87&70.26&76.71&58.54&83.27&73.10&63.88&59.67&74.10&72.63&70.59 \\
    \hline
    \hline
     nnUNet&\textbf{81.89}&83.10&78.71&60.08&75.21&82.06&\underline{66.21}&85.60&77.16&\underline{69.76}&\underline{70.29}&79.38&\textbf{79.93}&\underline{76.10} \\
    \hline
    +CBAM&81.77&84.13&78.95&58.66&75.55&\textbf{82.31}&64.70&85.91&76.79&69.51&69.88&78.40&79.37&75.84 \\
    \hline
    +AT&81.79&82.31&78.86&58.93&\underline{76.30}&82.02&65.07&\underline{86.27}&\underline{77.70}&69.66&69.98&78.68&\underline{79.90}&75.96 \\
    \hline
    +MAT&81.68&\underline{84.30}&79.11&57.98&75.64&81.86&65.15&85.93&77.50&68.99&69.39&78.77&79.42&75.83 \\
    \hline
    +GASA(Ours)&81.48&\textbf{84.88}&\underline{79.75}&\underline{60.28}&75.34&82.15&64.38&85.58&76.95&69.53&69.60&\underline{79.41}&78.99&76.03 \\
    \hline
    +GASA-L(Ours)&\underline{81.81}&83.78&\textbf{80.34}&\textbf{63.60}&\textbf{75.80}&\underline{82.27}&\textbf{66.96}&\textbf{87.04}&\textbf{78.14}&\textbf{69.90}&\textbf{71.70}&\textbf{80.25}&79.86&\textbf{77.04} \\
    \hline
  \end{tabular}}
  
  \caption{Comparison of segmentation performance on the BTCV dataset. 
  Abbreviations: Spl - spleen, Rkid - right kidney, Lkid - left kidney, Gall - gallbladder, Eso - esophagus, Liv - liver, Sto - stomach, Aor - aorta, IVC - inferior vena cava, Veins - portal and splenic veins, Pan - pancreas, Rag - right adrenal gland, Lag - left adrenal gland, Avg - average, Ours - GASA-UNet.}
  \label{tab:btcv}
\end{table*}
\begin{table*}
  \centering
\resizebox{0.9\textwidth}{!}{
  \begin{tabular}{@{}c|ccccccccccccc|c@{}}
    \hline
    \textbf{Dice}&Spl&Kid&Gal&Eso&Liv&Sto&Aor&Pca&Pan&AG&Duo&Bla&Ute&Avg\\
    \hline

    nnMamba&94.93&93.33&73.89&76.93&95.78&87.70&93.69&87.89&81.81&69.57&74.68&79.64&74.73&83.16\\
    \hline
    +GASA(Ours)&95.33&94.12&75.57&79.08&96.27&87.86&93.95&89.00&81.42&71.61&75.55&83.31&75.88&84.31\\
    \hline
    \hline
    UNETR&89.03&84.62&53.94&59.64&92.52&68.79&85.44&75.11&63.27&48.04&46.32&68.15&62.83&68.62\\
    \hline
    +GASA(Ours)&89.55&87.32&56.52&62.78&93.04&71.31&86.60&76.46&66.18&54.72&52.28&69.545&64.93&71.55\\
    \hline
    \hline
    V-Net&90.96&89.53&69.69&67.97&93.93&81.20&90.82&83.35&74.13&61.05&63.39&75.34&64.21&77.15\\
    \hline
    +GASA(Ours)&88.25&90.97&69.55&74.10&92.87&79.56&90.66&84.06&71.37&67.99&66.62&82.05&69.83&79.12\\
    \hline
    \hline
    nnUNet&97.14&96.71&83.90&\textbf{86.60}&97.56&92.44&95.64&92.28&88.72&79.70&\textbf{83.73}&\underline{90.08}&85.42&89.75\\
    \hline
    +CBAM&\underline{97.19}&\underline{96.72}&\underline{84.56}&\underline{86.50}&97.55&\underline{92.51}&95.65&92.30&88.64&79.76&83.56&90.06&\textbf{85.79}&89.81 \\
    \hline
    +AT&97.18&96.67&83.10&86.51&\underline{97.57}&92.41&\underline{95.66}&92.32&88.66&79.84&\underline{83.65}&89.08&85.11&89.62 \\
    \hline
    \hline
    Ours(Ours)&97.14&96.70&\textbf{84.69}&86.44&97.55&92.32&95.63&\underline{92.35}&\underline{88.74}&\underline{79.92}&83.57&90.06&\underline{85.75}&\underline{89.83}\\
    \hline
    Ours(L)&\textbf{97.43}&\textbf{96.76}&83.60&\underline{86.50}&\textbf{97.69}&\textbf{93.41}&\textbf{95.71}&\textbf{92.37}&\textbf{89.01}&\textbf{79.94}&83.53&\textbf{90.16}&85.29&\textbf{89.87}\\
    \hline
    \hline
    \textbf{NSD}&Spl&Kid&Gal&Eso&Liv&Sto&Aor&Pca&Pan&AG&Duo&Bla&Ute&Avg\\

    \hline
    nnMamba&85.17&87.08&63.39&67.69&77.84&69.28&90.33&78.66&67.04&73.23&61.26&61.53&52.24&73.00\\
    \hline
    +GASA(Ours)&86.79&88.59&66.49&72.95&79.60&70.59&91.28&81.03&66.56&76.31&62.97&65.39&54.12&75.17\\
    \hline
    \hline
    UNETR&68.72&70.39&39.31&50.04&66.12&38.71&71.43&53.99&40.57&48.74&32.72&40.58&34.28&51.65\\
    \hline
    +GASA(Ours)&70.81&73.55&42.12&53.10&68.21&41.04&73.86&56.46&43.87&56.78&37.20&40.83&36.86&55.00\\
    \hline
    \hline
 
    V-Net&75.56&79.19&56.31&57.58&71.94&55.74&84.02&68.71&53.59&62.69&46.58&53.01&37.75&62.97\\
    \hline
    +GASA(Ours)&71.44&78.40&59.94&68.62&65.11&54.10&84.64&71.37&51.20&73.67&52.72&63.40&45.66&66.16\\
    \hline
    \hline
    nnUNet&93.22&93.23&\textbf{81.96}&\underline{85.94}&87.94&82.27&94.32&\underline{87.90}&80.27&85.66&\underline{77.10}&81.93&73.10&85.58\\
    \hline
    +CBAM&\underline{93.36}&93.22&81.77&85.86&87.99&\underline{82.37}&\underline{94.42}&87.85&80.25&85.78&76.86&81.81&\underline{73.20}&85.58 \\
    \hline
    +AT&93.33&93.17&81.52&85.75&\underline{88.07}&82.15&94.38&87.88&80.24&85.85&76.91&81.98&73.09&85.56 \\
    \hline
    +GASA(Ours) &93.29&\underline{93.25}&81.89&85.75&88.00&82.17&94.34&\underline{87.90}&\underline{80.41}&\underline{85.95}&76.98&\underline{82.03}&\textbf{73.41}&\underline{85.64}\\
    \hline
    +GASA-L(Ours)&\textbf{93.89}&\textbf{93.33}&\underline{81.91}&\textbf{85.97}&\textbf{88.37}&\textbf{83.24}&\textbf{94.43}&\textbf{87.95}&\textbf{80.87}&\textbf{86.03}&\textbf{77.13}&\textbf{83.42}&73.15&\textbf{85.93}\\
    \hline
    
  \end{tabular}
  }

  \caption{Comparison of segmentation performance on the AMOS dataset.  Abbreviations: Kid - kidney, Pca - postcava, AG - adrenal gland, Duo - duodenum, Bla - bladder, Ute - uterus.}
  \label{tab:AMOS}
\end{table*}

\clearpage
\bibliography{aaai25}